\documentclass[preprint,authoryear,12pt]{elsarticle}

\usepackage[latin1]{inputenc} %Spanish language.
\usepackage{array}
\usepackage{multirow}
\usepackage{rotating}
\usepackage{graphicx}
\journal{}

\begin{document}

\begin{frontmatter}



\title{The use of blogs in the education field: A qualitative systematic review}


\author[label1]{Carlos R. del Blanco}
\author[label2]{Ivan García-Magariño}

\address[label1]{Universidad Politécnica de Madrid, ETSI Telecomunicación, GTI, Madrid, Spain, cda@gti.ssr.upm.es}
\address[label2]{Universidad Complutense de Madrid, Facultad de Informática, Madrid, Spain, ivan.gmg@fdi.ucm.es}

\begin{abstract}
Blogs have become one of the most successful tools of the Web 2.0 because of their ease of use and the availability of open platforms. They have quickly spread in the education field thanks to the many attractive qualities that have been attributed to them, such as collaboration, communication, enhancing of professional writing, and the improvement of information-gathering skills. However, many of the studies that have addressed this issue were not based on an empirical research, and therefore they are unreliable. On the other hand, the studies that do have conducted an empirical research have usually relied on participant self-reported data (surveys, interviews, and contents of blogs), which can significantly bias the positive results usually reported on the use of blogs. Another source of bias and inaccuracy in the reported results is that most of the studies lacked control group, i.e they do not follow an experimental design. The purpose of this review is to examine the current state of the studies related to the evaluation of the blog effects in the education field. The methods to select the studies and perform the corresponding analysis have followed a qualitative systematic approach. The selection has been restricted to empirical and peer-reviewed studies published between January 2011 and June 2013. The findings have been integrated and compared using the Grounded Theory, giving rise to a set of categories that have structured the results of the review.
\end{abstract}


\end{frontmatter}


\section{Introduction}

Blogs are nowadays one of the most popular tools of the Web 2.0 that have spread across many fields because of their ease of use and the availability of open and free platforms. Education is one of these fields where some researchers~\citep{RefWorks:206} have even suggested that a methodology based on blogs can be applied to different levels and disciplines. From an educational point of view, blogs have aroused high expectations because of the many desirable qualities that have been attributed to them, such as collaboration, communication, enhancing of professional writing, and improvement of information-gathering skills. Some researchers~\citep{RefWorks:207} have suggested that both the motivation of students and the quality of writings are improved due to the fact that the posted information is available to the general public. Blogs have also been considered as a tool that boosts the interactions among learners and between learners and instructors~\citep{RefWorks:208}.

The main problem with the existing works that have addressed the effects of blogs in education is the lack of empirical research. To shed light on this fact, \citet{RefWorks:101} performed a review to know the state of the empirical research on blogs in higher education and summarized the main findings. Although positive results were reported on the use of blogs to improve students' learning and the affective outcome, the authors argued that these results could be biased by the fact that most studies were exclusively based on self-report data. In addition, all the studies except one lacked a control group, i.e. they did not follow an experimental design. The conclusion was that more research had to be conducted to answer the questions on whether the use of blogs could increase the students' performance and the learning engagement.

In this review, the aim is to examine the evolution and current state of the studies that are concerned with the evaluation of the effects of blogs in the education field, serving as a natural extension and update of the review conducted by \citet{RefWorks:101}. Unlike the previous review, the present one is not limited to higher education, considering the research performed in other education levels, such as primary or secondary education. In this paper, the approach has been to perform a qualitative systematic review, also known as a qualitative evidence synthesis~\citep{RefWorks:211}. This type of review is featured by performing a thorough and exhaustive selection of empirical and peer-reviewed studies, which have been analyzed using the Grounded Theory~\citep{RefWorks:191}, also called constant-comparative method. As a result of the analytical review, a set of categories or dimensions have been obtained, which in turn have structured the main findings of the review.

The rest of the paper is organized as follows. Section~\ref{sec:lit} summarizes the main characteristics of blogs, and presents a discussion of the reviews focused on the use of blogs in the education field. Section~\ref{sec:sources} describes the procedure to select the studies that are focused on the effects of blogs in education, while section~\ref{sec:method} explains the analytical method used in the review process. Section~\ref{sec:findings} reports the main findings resulting from the review process, which are structured in categories. Finally, conclusions are drawn in section~\ref{sec:conc}.

\section{Literature review}
\label{sec:lit}

Web 2.0 tools, also known as Web-based collaborationware, have spread with great success in a broad range of professional activities and educational services in the last decade. In general, they offer many interesting and useful features such as powerful information sharing, good experience of collaboration, rapidity of deployment, and ease of use. Blogs or weblogs are one the most popular Web 2.0 tools, which have been used not only for the professional activity and education of students, but also for the continue update of knowledge and competencies of both professionals and students. 

Blogs allow the creation of Web sites that are characterized by the following features: (1) reverse chronological order, (2) archival of posts, (3) hypermedia content, and (4) ease of use. The first feature, reverse chronological order, refers to the fact that a blog contains dated entries/posts that are displayed in reverse chronological order. Therefore, the last post is arranged at the top of the page, whereas older posts appear further down. This feature allows readers to access to the most updated information of the blog. The second feature, archival of posts, entails that a blog only displays a fixed number of recent posts, while the old ones are automatically archived. This allows to focus on the last and more recent posts, avoiding that the number of displayed posts grow indefinitely. Archived posts can be usually accessed by different methods: using a permanent link (called permalink) making use of an internal search engine (if available), or browsing over the directory of dates in which the posts were created. The permalink facilitates the linking to content within the own blog and from external sites. The third feature, hypermedia content, is related to the fact that posts in a blog can contain multitude of links to resources of different kinds: text, audio, video, other sites, and blogs. Nowadays, it is also common that blogs have the capacity to embed in the own posts multimedia content and little applications that offer a great range of functionalities. The fourth and last feature, ease of use, has been one of the keys for the popularity and fast spreading of the blogs. A blog can be created and managed by a user without any specific knowledge in programming and databases. It is almost at the same level that writing a document with a simple text editor.

Other potential characteristics that a blog can have are: the inclusion of a search engine to easily and quickly locate specific information spread in the collection of posts; the possibility that multiple users can collaborate in the same blog, who can have different kinds of roles to improve the management and creation of content; and the posting of comments for third-party users, fostering the dialog and feedback with the internet community.

Regarding the types of blogs, \citet{RefWorks:115} made a classification according to their genre, finding three types: (1) filters (the blog is used to comment the content of other websites), (2) personal journals (the blog content is focused on the thoughts and life of the author), and (3) knowledge logs (blogs that gather information, references and conclusions about a specific knowledge domain). In addition, they found that the the predominant type of blogs were the personal journals.

Although several reviews have arisen about the use of Web 2.0 tools for educational purposes in the last years, few of them are focused on blogs. The researchers \citet{RefWorks:212} conducted a review on the use of Web 2.0 tools in higher education from 2007-2009. They found that blogs were one of the five tools most commonly discussed in the literature. In addition, they indicated that the research evidence of the effectiveness of these tools was unclear to enhance teaching and learning. On the other hand, instructors and students reported a positive feedback about the aforementioned Web 2.0 tools. \citet{RefWorks:214} conducted another review about the role of Web 2.0 tools in education, which included blogs. Her findings indicated that the use of Web 2.0 for teaching purposes could foster a more collaborative approach to study. In the field of cardiology, \citet{RefWorks:100} described the most useful applications of blogs and other Web 2.0 tools for the education and knowledge update. More relevant is the review conducted by \citet{RefWorks:101} in 2010, since it is the only one totally focused on blogs. This review assesses the state of the empirical research on blogs in higher education, and summarizes the main findings. Their review indicated that the effects of blogging in the dimension of the performance outcome could help in student learning and develop thinking skills. The review also examined the blog effects on the affective outcome of students, such as their attitudes and satisfaction. They found that most students have a positive perception towards the use of blogs due to their ease of use, and that they would like to have the opportunity to perform blogging activities in more courses. However, authors also warned that almost all the studies selected for the review relied on participant self-reported data (surveys, interviews, and contents of blogs), and therefore they can have a noticeable bias in their results because the involved participants tend to tune the answers towards what would be socially desirable answers~\citep{RefWorks:210}. In addition, only one of these studies~\citep{RefWorks:209} followed an experimental design, dividing the sampled population into a control and a treatment group, which is a desirable study feature to provide more robust and less biased results. On the whole, this review concluded that the questions on whether the use of blogs can increase the performance and affective outcome are still unresolved.

The purpose of the presented review is to examine the evolution and current state of the studies that are concerned with the evaluation of the effects of blogs in the education field, serving as a natural extension and update of the review conducted by \citet{RefWorks:101}. One substantial difference with the previous review is that the current one is not limited to higher education, considering studies that involve other education levels such as primary and secondary education. The methods to select the studies and perform the corresponding analysis have followed a qualitative systematic approach, also known as a qualitative evidence synthesis~\citep{RefWorks:211}. According to this type of review, a thorough and exhaustive selection of empirical and peer-reviewed studies has been performed. The findings of the selected studies have been integrated and compared using the Grounded Theory~\citep{RefWorks:191}, also called constant-comparative method, to finally derive a set of categories that have structured the findings of the review.

\section{Sources of data}
\label{sec:sources}

The search of scientific papers focused on the use of blogs in the education field has been carried out in the following meta-search engines: ISI Web of Knowledge and Ingenio (from the \emph{Universidad Politécnica de Madrid}). These meta-search engines have access to a broad range of databases, including the most relevant ones, such as Elsevier, Scopus, Web of Science, Education Resources Information Center (ERIC), Academic Search Premier, Medical Literature Analysis and Retrieval System Online (MEDLINE), PubMed, PsycINFO, and Educational Search Complete.

The terms used to perform the searches have been: blogs, weblogs, edublogs, and education. The searches have been restricted to dates between January 2011 and June 2013 to evaluate the evolution of the use of blogs in education with respect to the work of \citet{RefWorks:101}. The searches were also restricted to those that had been peer-reviewed with the purpose of filtering non-relevant or non-significant works. Lastly, only English-written papers have been considered.

As of 10 June 2013, a total of 882 articles were found. Out of these, 47 were identified for this review. The main reason that led to discard a work from the final selection were that it was non-empirical or not concerned with educational purposes. In addition, those studies that combined the use of blogs with other web 2.0 tools were also discarded with the purpose to assess the actual efficiency of blogs, and avoiding uncontrolled relationships among several web technologies. A fundamental difference with regard to the review of \citet{RefWorks:101} is that the present review is not limited to higher education, considering all levels of education: from primary up to doctoral education.

\section{Methodology}
\label{sec:method}

One of the questions to be answered in this review was if there would have been enough experimental studies to perform a meta-analysis. A meta-analysis is a convenient technique to summarize the accumulated state of knowledge about a specific topic of interest, which statistically combines quantitative studies results to provide a more precise analysis of their results. The factors that lead to a successful meta-analysis study depends on the research design of the involved studies, the characteristic of the population under review, and the homogeneity/heterogeneity of the results. For the ongoing review about the use of blogs in education, only 6 studies out of 47 were based on an experimental design with a control group and a treatment group. On the other hand, the population of the most of the selected studies are too small to be statistically significant. These both facts preclude the application of meta-analytical techniques. For this reason, an alternative approach for the development of the review has been adopted, called qualitative systematic review or qualitative evidence synthesis~\citep{RefWorks:211}. This is a method for integrating and comparing the findings from qualitative studies that makes use of the Grounded Theory~\citep{RefWorks:191}, also called constant-comparative method. The collection and analysis of data is performed simultaneously, where the basic unit of data is (in our context) each individual study described in an article. The analysis of the collected studies involves the following stages: creation of analytic codes inferred from the own studies and not by pre-existing conceptualizations, grouping of codes into similar concepts, and formation of categories from concepts. The process by which the categories are obtained is iterative: new concepts are compared with existing categories for grouping, and if one concept does not fit any of them appropriately, a new category is created. This process continues up to achieving the category saturation. The obtained categories give rise to the findings of the review, summarizing the predominant topics, presenting the limitations of the current research, and proposing new directions for future research.

\section{Findings}
\label{sec:findings}

As a result of the qualitative systematic review, the following five categories have been identified: (1) effects of blogging on reflective practice and learning, (2) effects of blogging on learning engagement, (3) effects of blogging on collaborative learning and students' interaction, (4) effects of blogging on social support, and (5) effects of blogging on the development of specific skills. Table \ref{tab:1} shows to what categories the studies belong. Notice that one study can belong to more than one category, since it can have addressed several topics. In the next subsections, the different categories are presented along with the involved studies. To avoid redundant information in the cases that one study is associated to several categories, it is fully described in the category that is considered as the most important one (because of extension or quality of the results), whereas only the key information is presented in the other categories.

\begin{sidewaystable}
%\begin{table}[h]
\resizebox{20cm}{!}{
\begin{tabular}{|l|l|l|c|c|c|c|c|c|}
\hline 
\multirow{5}{*}{\textbf{Study}} & \multirow{5}{*}{\textbf{Participants}} & \multirow{5}{*}{\textbf{Type of data}} & \multirow{5}{*}{\textbf{Control group}} & \multicolumn{5}{c|}{\textbf{Categories}}\tabularnewline
\cline{5-9} 
 &  &  &  & \textbf{Reflective} & \textbf{Learning} & \textbf{Collaborative} & \textbf{Social} & \textbf{Development}\tabularnewline
 &  &  &  & \textbf{practice} & \textbf{engagement} & \textbf{learning and} & \textbf{support} & \textbf{of specific}\tabularnewline
 &  &  &  & \textbf{and learning} &  & \textbf{students'} & \multirow{2}{*}{} & \textbf{skills}\tabularnewline
 &  &  &  &  &  & \textbf{interaction} &  & \tabularnewline
\hline 
Alqudsi-Ghabra and  & \multirow{2}{*}{14 undergraduate students} & \multirow{2}{*}{Survey} & \multirow{2}{*}{No} & \multirow{2}{*}{X} & \multirow{2}{*}{} & \multirow{2}{*}{} & \multirow{2}{*}{X} & \multirow{2}{*}{}\tabularnewline
Al-Bahrani (2012) &  &  &  &  &  &  &  & \tabularnewline
\hline 
Alterman and Larusson (2013) & 25 undergraduate students & Contents of blog, survey & No &  &  & X &  & \tabularnewline
\hline 
Angelaina and Jimoyiannis (2012) & 21 secondary students & Contents of blog & No &  & X & X &  & \tabularnewline
\hline 
Askim et al. (2011) & 68 undergraduate\ students & Interview, survey & No &  & X &  &  & \tabularnewline
\hline 
\multirow{2}{*}{Bartholomew et al. (2012)} & Undergraduate students  & Contents of blog,  & \multirow{2}{*}{No} & \multirow{2}{*}{X} & \multirow{2}{*}{X} & \multirow{2}{*}{} & \multirow{2}{*}{} & \multirow{2}{*}{}\tabularnewline
 & (number not specified) & participant observation &  &  &  &  &  & \tabularnewline
\hline 
Cain and Giraud (2012) & 17 postgraduate students & Contents of blog & No & X &  &  &  & \tabularnewline
\hline 
\multirow{2}{*}{Cameron (2011)} & \multirow{2}{*}{170 undergraduate students} & Contents of blog, survey, assignment,  & \multirow{2}{*}{No} & \multirow{2}{*}{X} & \multirow{2}{*}{X} & \multirow{2}{*}{} & \multirow{2}{*}{} & \multirow{2}{*}{}\tabularnewline
 &  & academic test, final exam &  &  &  &  &  & \tabularnewline
\hline 
\multirow{2}{*}{Chen (2012)} & \multirow{2}{*}{67 undergraduate students} & Survey, document analysis,  & \multirow{2}{*}{No} & \multirow{2}{*}{} & \multirow{2}{*}{} & \multirow{2}{*}{} & \multirow{2}{*}{} & \multirow{2}{*}{X}\tabularnewline
 &  & participant observation &  &  &  &  &  & \tabularnewline
\hline 
Chhabra and Sharma (2011) & 133 undergraduate students & Survey, academic test & Yes & X & X &  &  & \tabularnewline
\hline 
Chu et al. (2012) & 81 undergraduate students & Contents of blog, interview & No & X & X &  & X & \tabularnewline
\hline 
\multirow{2}{*}{Churchill (2011)} & \multirow{2}{*}{26 postgraduate students} & Interview, survey,  & \multirow{2}{*}{No} & \multirow{2}{*}{X} & \multirow{2}{*}{} & \multirow{2}{*}{} & \multirow{2}{*}{} & \multirow{2}{*}{}\tabularnewline
 &  & final exam, participant observation &  &  &  &  &  & \tabularnewline
\hline 
Deed and Edwards (2011) & 400 undergraduate students & Contents of blog, surveys & No & X &  &  &  & \tabularnewline
\hline 
Deng and Yuen (2011) & 37 undergraduate students & Contents of blog, interview, survey & No & X &  & X & X & \tabularnewline
\hline 
Fessakis et al. (2013) & 147 undergraduate students & Contents of blog, survey & No &  &  & X &  & \tabularnewline
\hline 
Garcia-Sabater et al. (2011) & 16 postgraduate students & Contents of blog & No & X &  &  &  & \tabularnewline
\hline 
García-Sánchez and  & \multirow{2}{*}{87 postgraduate students } & Contents of blog, surveys,  & \multirow{2}{*}{No} & \multirow{2}{*}{} & \multirow{2}{*}{} & \multirow{2}{*}{X} & \multirow{2}{*}{} & \multirow{2}{*}{X}\tabularnewline
Rojas-Lizana (2012) &  & participant observation &  &  &  &  &  & \tabularnewline
\hline 
Goktas and Demirel (2012) & 339 undergraduate students & Interview, survey & No & X & X &  &  & X\tabularnewline
\hline 
Hamad Aljumah (2011) & 35 undergraduate students & Survey & No & X &  &  &  & X\tabularnewline
\hline 
Harland and Wondra (2011) & 67 undergraduate students & Contents of blog and hard copy paper & Yes & X &  &  &  & \tabularnewline
\hline 
Hodgson and Wong (2011) & 524 undergraduate students & Contents of blog, interview, survey & No &  &  & X &  & X\tabularnewline
\hline 
\multirow{2}{*}{Hsu and Wang (2011)} & \multirow{2}{*}{149 undergraduate students} & Survey, assignment,  & \multirow{2}{*}{Yes} & \multirow{2}{*}{X} & \multirow{2}{*}{X} & \multirow{2}{*}{} & \multirow{2}{*}{X} & \multirow{2}{*}{}\tabularnewline
 &  & academic test, final exam &  &  &  &  &  & \tabularnewline
\hline 
\multirow{2}{*}{Hume (2012)} & \multirow{2}{*}{176 postgraduate students} & Contents of blog, interview, survey,  & \multirow{2}{*}{No} & \multirow{2}{*}{X} & \multirow{2}{*}{} & \multirow{2}{*}{} & \multirow{2}{*}{} & \multirow{2}{*}{}\tabularnewline
 &  & assignment &  &  &  &  &  & \tabularnewline
\hline 
Hungerford-Kresser et al. (2011) & 120 undergraduate students & Contents of blog, interview & Yes & X &  &  &  & \tabularnewline
\hline 
\multirow{2}{*}{Jara (2012)} & Primary students & \multirow{2}{*}{Contents of blog, assignment} & \multirow{2}{*}{No} & \multirow{2}{*}{} & \multirow{2}{*}{X} & \multirow{2}{*}{} & \multirow{2}{*}{} & \multirow{2}{*}{X}\tabularnewline
 &  (number not specified) &  &  &  &  &  &  & \tabularnewline
\hline 
Jimoyiannis (2012) & 21 secondary students & Contents of blog & No & X & X & X & X & \tabularnewline
\hline 
Kang (2011) & 24 undergraduate students & Contents of blog & No &  &  & X &  & \tabularnewline
\hline 
Maheridou (2012) & 20 postgraduate students & Contents of blog & No &  &  & X & X & \tabularnewline
\hline 
\multirow{2}{*}{Manfra and Lee (2012)} & Secondary students  & Contents of blog, interview, & \multirow{2}{*}{No} & \multirow{2}{*}{} & \multirow{2}{*}{X} & \multirow{2}{*}{} & \multirow{2}{*}{} & \multirow{2}{*}{}\tabularnewline
 & (number not specified) &  participant observation &  &  &  &  &  & \tabularnewline
\hline 
Marques et al. (2013) & 26 undergraduate students & Contents of blog, interview, survey & No &  & X & X &  & \tabularnewline
\hline 
\multirow{2}{*}{McGrail and Davis (2011)} & \multirow{2}{*}{16 primary students} & Contents of blog, interview,  & \multirow{2}{*}{No} & \multirow{2}{*}{} & \multirow{2}{*}{} & \multirow{2}{*}{} & \multirow{2}{*}{X} & \multirow{2}{*}{X}\tabularnewline
 &  & participant observation &  &  &  &  &  & \tabularnewline
\hline 
\multirow{2}{*}{Mohd Hashim (2012)} & Postgraduate students  & Contents of blog, survey, interview,  & \multirow{2}{*}{No} & \multirow{2}{*}{X} & \multirow{2}{*}{} & \multirow{2}{*}{} & \multirow{2}{*}{} & \multirow{2}{*}{}\tabularnewline
\cline{3-3} 
 & (number not specified) & participant observation &  &  &  &  &  & \tabularnewline
\hline 
Nehme (2011) & 17 undergraduate students & Contents of blog, survey & No & X & X &  &  & \tabularnewline
\hline
\multirow{2}{*}{Olofsson et al. (2011)} & 23 undergraduate and  & \multirow{2}{*}{Contents of blog} & \multirow{2}{*}{No} & \multirow{2}{*}{X} & \multirow{2}{*}{} & \multirow{2}{*}{X} & \multirow{2}{*}{} & \multirow{2}{*}{}\tabularnewline
 & graduate students &  &  &  &  &  &  & \tabularnewline
\hline 
Osman and Koh (2013) & 65 postgraduate students & Contents of blog & No & X &  &  &  & \tabularnewline
\hline 
Papastergiou et al. (2011) & 70 undergraduate students & Survey, academic test & Yes & X & X &  &  & X\tabularnewline
\hline 
Pinilla et al. (2013) & 33 undergraduate students & Contents of blog & No & X &  &  & X & \tabularnewline
\hline 
Reupert and Dalgarno (2011) & 91 undergraduate students & Contents of blog, interview & No & X &  & X & X & \tabularnewline
\hline 
Robertson (2011) & 113 undergraduate students & Contents of blog & No &  &  & X & X & \tabularnewline
\hline 
\multirow{2}{*}{Rojas (2011)} & \multirow{2}{*}{43 secondary students} & Contents of blog, survey, interview,  & \multirow{2}{*}{No} & \multirow{2}{*}{} & \multirow{2}{*}{X} & \multirow{2}{*}{} & \multirow{2}{*}{} & \multirow{2}{*}{}\tabularnewline
 &  & participant observation &  &  &  &  &  & \tabularnewline
\hline 
Stevens and Brown (2011) & 2 postgraduate students & Contents of blog, interview, survey & No & X &  &  &  & \tabularnewline
\hline 
Sun and Chang (2012) & 7 postgraduate students & Contents of blog & No & X &  &  & X & X\tabularnewline
\hline 
Suryani et al. (2012) & 37 undergraduate students & Survey, academic test & No &  &  &  &  & X\tabularnewline
\hline 
Tang (2012) & 49 undergraduate students & Contents of blog & No & X &  &  &  & \tabularnewline
\hline 
Top (2012) & 50 undergraduate students & Survey & No & X &  &  &  & X\tabularnewline
\hline 
Vurdien (2013) & 11 language school students & Contents of blog, survey & No &  &  & X &  & X\tabularnewline
\hline 
Wood (2012) & 14 undergraduate students & Contents of blog, interview, survey & No & X &  &  &  & \tabularnewline
\hline 
\multirow{2}{*}{Yang and Chang (2012)} & 154 undergraduate and  & \multirow{2}{*}{Contents of blog, survey} & \multirow{2}{*}{Yes} & \multirow{2}{*}{} & \multirow{2}{*}{X} & \multirow{2}{*}{} & \multirow{2}{*}{X} & \multirow{2}{*}{}\tabularnewline
 & graduate students &  &  &  &  &  &  & \tabularnewline
\hline 
\end{tabular}
}
\caption{Summary of selected studies and their attributed categories.}
\label{tab:1}
%\end{table}[h]
\end{sidewaystable}
%---------------------------------------------------------------------------------------------

%---------------------------------------------------------------------------------------------
\subsection{Effects of blogging on reflective practice and learning}

A common interest among the researchers was to examine if the use of blogs for educational purposes could foster the reflective practice in order to improve the student learning. The concept of reflection has been defined by \citet{RefWorks:202} as ``an important human activity in which people recapture their experience, think about it, mull it over and evaluate it. It is this working with experience that is important in learning.'' The reflective practice is also related to the concepts of deep thinking, reflective learning, critical thinking, deeper understanding, knowledge construction, and experiential learning. These terms arise from the several models of reflection that exist in literature, such as those of \citet{RefWorks:194}, \citet{RefWorks:195}, \citet{RefWorks:197}, \citet{RefWorks:198}, \citet{RefWorks:199}, and \citet{RefWorks:200}.

The majority of the analyzed studies (24 out of 29) used data collecting methods based on participant self-reported data, such as surveys, interviews, and content of the blogs. Besides, only 5 out 29 followed an experimental design, i.e. they had treatment and control groups.  

Regarding the studies that only used self-reported data and lacked control group, the results obtained by 16 out of 19 studies showed that blogging activities can foster reflective practices, which in turn help in the development of the students' learning. For example, the study conducted by \citet{RefWorks:119} focused on the application of an analysis framework for the evaluation of the learning achieved from blogging activities. The framework was based on the Community of Inquiry and the Social Network Analysis models. A total of 21 secondary students from an Information and Communication Technology (ICT) course participated in the study. They were asked to post entries in a class blog to reflect upon and discuss about course contents. The analysis of the content of the blogs indicated that students achieved the integration of ideas and the construction of meaning. \citet{RefWorks:125} interviewed 81 undergraduate students from information management and nursing to examine the process of learning and the student perceptions in the practice of blogging during an internship. Participants were asked to create individual blogs to share their experiences and useful learning resources. The study found that students perceived blogs as a useful tool for knowledge construction, reflection, and problem solving. \citet{RefWorks:132} surveyed 50 pre-service teachers to examine the perceived learning, the perceived collaborative learning, and the sense of community in ICT courses that incorporated blogs. Participants were required to collaboratively write posts in group blogs to carry out an instructional project. The study revealed that students had positive perceptions about their learning and collaborative learning through blogs. \citet{RefWorks:135} surveyed and interviewed 14 pre-service teachers of geography to assess the convenience of using blogs to develop students' understanding, and to prepare them for teaching practice. Pre-service teachers were asked to work in pairs to create a blog focused on geographical concepts that were particularly difficult to understand. The study found that blogs acted as liminal spaces where students underwent deep changes in understanding as a result of their learning and experiences. In other words, blogs allowed students to start their transition from trainees to competent teachers. \citet{RefWorks:138} surveyed 17 undergraduate students from two universities to evaluate the impact of blogging on students' learning, deep understanding, and student motivation. Students were asked to post entries in a class blog about real-life applications of linear algebra, and write comments on other peer posts. The study found that students constructed new knowledge as a result of the blogging activity and benefited from the cooperative learning experience. \citet{RefWorks:148} conducted a study with the participation of 23 graduate and undergraduate students of an online ICT course to examine the possibilities of blogs to support reflective peer-to-peer learning and formative assessment practice. Students were required to post entries in individual blogs, and respond to comments of their peers. Students had to reflect upon the course literature and lectures offered in a podcast format. The study found that blogging activities could improve formative assessment skills, and develop connections between new knowledge and prior knowledge. \citet{Osman2013} examined if the use blog could promote the critical thinking and the theory-practice linkage within a Master-Bussiness-Administration course with 65 students. Students were required to post entries and make comments on peers contributions in group blogs. The findings positively pointed to the potential of blogs to foster the learning and student's reflective practices. \citet{Pinilla2013} conducted a qualitative study to examine the use of blogs by medical students and to investigate the students' experiences and impressions. The contents of 33 blogs were analyzed, which were all authored by medical students. The study found that blogs are useful tools for reflecting on the experiences during medical school. \citet{Alqudsi-Ghabra2012} conducted a study with 14 undergraduate students to examine the use of blogs in higher education by means of a comparative analysis of two different blogging experiences. Half of the students were required to participate in structured and course-centered blog, while the other half voluntarily participated in a student-centered blog. Results indicated that both types of blogging activities supplement each other, leading to high-order thinking skills. \citet{Tang2012} examined the students' reflective process in the context of blogging. 49 pre-service teachers participated in the study writing reflections and comments in a common blog about their experiences in the teaching practicum. The study reported that the analysis of the frequency of blog entries suggested that students were active in the reflective activity. \citet{Cain2012} conducted a study with 17 postgraduate students to examine the critical thinking skills in the context of leadership using structured and unstructured blogs. Students were required to write in personal blogs as part of course grade. Results indicated that self-regulation and explanation were the two most consistent critical thinking skills among most students. In addition, students that participated in structured blogs were more thought evoking than those that wrote in unstructured blogs. \citet{HamadAljumah2011} investigated students' attitudes toward the use of blog to enhance the writing skills of a foreign language. Learners were asked to write one entry per week, and commenting on other classmates' blogs. Results indicated that students have a favorable perception towards the use of blogs for their writing assignments, and they positively assessed the feedback and interaction from classmates and teachers. \citet{Garcia-Sabater2011} conducted a study with 16 master students to analyze the utility of blogs in the teaching and learning of a Inventory Management course. Participants were required to frequently post entries about topics explained in the course and other information relevant to the students. Results indicated that students considered useful the blogging activities for the acquisition of knowledge and for the customization of the learning experience. Also, it was found that it is a mistake to assume that students master the use of web 2.0 technologies. Other researchers that also reported positive results in this line were \citet{RefWorks:123} for the acquisition of ICT competencies, \citet{RefWorks:128} for the improvement of academic writing skills, and \citet{RefWorks:147} in the context of teaching education.

On the other hand, three studies also relying on self-reported data and lacking control group informed mixed or non-concluding results. For example, \citet{RefWorks:152} surveyed and interviewed 2 master students (and teachers) in the context of an ICT course to examine the effects of blogs in promoting the use of information and communication technologies. Students were required to post reflective entries in both a class blog and an individual blog about their current technology proficiency and perceptions of their literacy instruction. They were also asked to respond to postings of others and request for advice from classmates. Mixed results were reported on the effects and perceptions of blogs. Blogs showed potential to promote ICT literacy and thoughtful reflection, however only a limited number of posts demonstrated such level of deep thinking. In addition, the students' perceptions about the support of blogs in their learning were varied and non-conclusive. \citet{Deed2011} conducted a study with 400 undergraduate students to evaluate the convenience of using blogs to promote the students' active learning and support higher-order thinking processes. Participants were required to write posts in groups to discuss and debate topics proposed in the course. Results indicated that the main goal of critical construction of knowledge was not totally achieved, although the students did seriously try to complete the assigned tasks. This fact was attributed to the devoid of guidance in the use of blogs for academic purposes, since it was wrongly assumed that students had enough expertise with the web 2.0 technologies. \citet{RefWorks:201} also reported that, although blog contents of the participants showed some evidence of reflection, the main use of blogs were to share tips and hints.

Five studies collected other types of data such as academic tests, examinations, and participant observation, in addition to surveys, interviews, and content of blogs. But, they still lacked control group. Four out of five studies found that the use of blog can promote the development of reflective activities and improve the students' learning. \citet{RefWorks:150} interviewed and surveyed 26 master students (and also school teachers) taking ICT courses with the purpose of developing a set of recommendations in the use of blogs for educational applications. In addition to self-reported data, end-of-course evaluation and participant observation were used in the evaluation of the effects of blogs. Participants were asked to write reflections about the learning tasks performed in the course in individual blogs, and comment on the contributions of each other. Moreover, a class blog was available so that students could access course materials. The study found that blogs had the potential to foster and improve the students' learning provided that blog usage is appropriately managed by the instructor. The following blog-related activities were recommended for maximizing the blog learning and teaching potential: reading blogs of peers, receiving comments from classmates, and reviewing completed tasks of other students along with the related existing feedback. \citet{Hume2012} conducted a study with 176 postgraduate students to assess whether educational blogs can enhance critical thinking and self-reflection of both students and lecturers. Participants were asked to post and comment their experiences about the course in a classroom group. Findings shown an increase in the students' overall performance, measured by an expert judgment panel. In addition, it is suggested that educators can benefit from blogging experiences to learn and teach more effectively. \citet{Bartholomew2012} conducted a study with undergraduate students to evaluate and develop blog management strategies to improve the students' learning and engagement. Participant observation along with the contents of blogs was used as data collection. Part of the students were required to post writing assignments, while others had to use the blog for discussion purposes. Results suggested that blogs encourage the student's learning, highlighting the commentary functionality that fosters the self-reflection and the participation. \citet{MohdHashim2012} examined the students' perception about the usefulness of blogs in a class of Masters in Technical and Vocational Education. Participants were encouraged to post their reflections on lessons and provide feedback to their classmates' comments. In addition to data collecting methods based on participant self-reported data, the method of participant observation was used. The study found that students have a positive perception about the use of blogs for educational purposes. However, they also demanded prior knowledge and training about blogging and computers skills. Unlike the previous studies, the one conducted by \citet{RefWorks:134} had different results. He examined whether the use of blogs could improve student learning performance, using as collected data academic tests, examinations, and other forms of self-reported data. A total of 170 students of Economics participated in the study using a class blog to start discussions and contribute to other discussions. Although students generally perceived positive feelings regarding the blog usage, the results indicated that student performance in tests and examinations was not correlated with blog participation once that student ability was controlled.

Five studies adopted an experimental design, dividing the participants in control and treatment groups. Two studies relied only on self-reported data, and the other three used other types of data such as academic tests, assignments, and final exams. Mixed results were reported about the correlation between the use of blogs and the acquired learning of students. This fact is quite significant since the results of these studies do not share the trend of the other studies that lacked an experimental design. These observations are derived from the following studies. Three studies reported no significant increase in students' knowledge or leaning. \citet{RefWorks:133} surveyed 120 pre-service teachers to explore the effects of blogging on facilitating classroom communities of practice in content-area literacy courses. The students belonging to the experimental group were required to post opinions and comments of textbook readings in a class blog, while the control group did not use any kind of blog. The study found no significant difference in students' perceptions of the acquired learning between the treatment and experimental groups. Moreover, students assessed blogs as the least important learning tool in the course. Researchers conjectured that one plausible reason for this result was the lack of direct instruction for performing blogging activities, since they incorrectly assumed that students would enjoy blogging. \citet{RefWorks:137} surveyed and evaluated (through academic tests) 70 undergraduate students to evaluate the potential of multimedia blogging for physical education, a discipline that is not heavily based on written discourse. More specifically, the study had the purpose to examine the effects of a multimedia class blog in the acquisition of knowledge of a selected sport in particular, and in the acquisition of ICT skills in general. The treatment group of 35 students was required to participate in a common blog creating multimedia posts, and reading the feedback from instructors, peers and an external expert. The control group of other 35 students did not perform any blogging activity, only reading an equivalent multimedia website. Findings indicated that blogging activity had no significant increase in students' knowledge, although students' perception was positive. \citet{Hsu2011} conducted a study with 149 undergraduate students to examine whether the use of blogs could improve the students' reading level. The treatment group, composed by 40 participants, was required to write weekly assignments in the form of blog entries and make comments on colleagues' posts. The other students, who formed part of the treatment group, submitted their assignments in the form of hard copies. Surveys, academic tests, and final course grades were analyzed. Results indicated that blogging did not represent an advantage that improves the students' reading performance. Other two studies reported positive results about the use on blogs. \citet{RefWorks:142} conducted an experimental study with 67 pre-service teachers of a curriculum and instruction course to examine the level of reflection achieved in their writing as effect of the use of blogs. Participants had to complete a writing assignment related to their clinical experiences. The students of the treatment group were required to write posts about classroom observations in individual blogs, and responding to classmate's comments. On the other hand, the students of the control group had to write an end-of-the-semester reflective paper. The study found that the writing of pre-service teachers who had used blogs showed higher levels of reflection in comparison with those students who had written papers. In addition, the study did not find any relationship between the depth of reflection and the peer/instructor feedback. Blog based writing had also the advantage of being 1,108 words shorter on average than the end-of-the-semester papers, which had 3,162 words on average (i.e. a 35.0\% reduction). \citet{Chhabra2011} surveyed and evaluated (using academic tests) 133 undergraduate engineering students to evaluate the potential of blogs to increase the academic performance in the context of problem based learning (PBL). Students within the treatment group were required to post problems about their assignments, and also to answer to the other students' problems. On the other hand, students within the control group did not perform any blogging activity. Results indicated that the academic performance of the group using the blog was better than the other group that did not use it.

% IVÁN (14 feb 2013): Carlos, el anterior dato de "1,000 words shorter on average" aporta poca o casi ninguna información, dado que depende mucho de la extensión media de los trabajos. Es decir, no es lo mismo que sea 1,000 palabras más cortas sobre una extensión media de 3,000 palabras que sobre una extensión media de 20,000 palabras. Por tanto, mira a ver si puedes extraer del artículo original la extensión media de los trabajos para que el lector pueda tener una información más adecuada. Otra opción es mencionar el tanto por ciento que son más cortos en media atendiendo al número de palabras. El tanto por ciento es mucho más indicativo en este caso que hablar únicamente de palabras más cortas sin indicar la extensión media.
% CARLOS R. (19 feb 2013): ok, hecho.
% IVÁN (20 feb 2013): He accedido al artículo original, y he puesto cifras más precisas en nuestro artículo para ser rigurosos, indicando como detalle adicional el tanto por ciento de reducción, para que el lector se pueda hacer una idea de la reducción sin tener que hacer los cálculos. Las cifras tan redondas que había en el artículo podrían haber levantado la sospecha en algún revisor.
%---------------------------------------------------------------------------------------------

%---------------------------------------------------------------------------------------------
\subsection{Effects of blogging on learning engagement}

The community of researchers was also interested in students' perceptions about the use of blog and its potential to increase the student engagement in learning tasks.

Most of the selected studies (11 out of 16) have exclusively relied on self-reported data such as on surveys, interviews, and contents of the blogs. In addition, only four out of 16 followed an experimental design.  

The studies that were based on self-reported data and lacked control group concluded that students had positive feelings towards the use of blogs and a higher interest in learning activities. For instance, \citet{RefWorks:141} surveyed and interviewed 68 undergraduates of an ICT course with the purpose of exploring the students' perceptions about blog usage for supplementary purposes in courses. Students were required to write posts in individual blogs to reflect upon course subjects. In addition, they had to discuss and respond questions posed by the instructor in a class blog. The study found that students who had engaged in the blog activities had higher levels of achievement and interest in the course. Moreover, the findings showed a significant difference in the level of achievement in favor of female students. On the other hand, the gender was not a determinant feature regarding the students' socialization and interest in courses, or the duration of their preparation for exams. \cite{Manfra2012} conducted a study with secondary students to explore the affordances and limitations of a group blog to engage students in the context of a history classroom. Participants were required to answer questions posted in the class blog, and also to read and respond to one another's contributions. Results indicated the students were able to engage in the educational activities through the use of blogs, especially when a single source of information and proper guidance were considered. This fact was attributed to the affordances provided by blogs, such as the student interaction, new kinds of interaction with the multimedia content, and the students' ability to direct their own learning. \cite{RefWorks:134} examined whether blogs could increase student engagement in the application of economics to real world situations, and reported that blogs indeed allowed students to be more engaged and aware of current economic issues. \citet{RefWorks:123} explored the use of blogs to assess their effects on the promotion of students' active engagement and the students' perceptions in blog-enhanced ICT courses. The study found that students positively changed their ICT perceptions. The study conducted by \citet{RefWorks:125} found that students from information management and nursing perceived the use of blogs as positive, and engaged in the blogging activity. Furthermore, positive perceptions were not influenced by the discipline of study, the frequency of use, or the blogging platform. \citet{RefWorks:119} reported that students from an ICT course demonstrated interest for the study and willingness in the participation of blogging activities. \citet{RefWorks:138} examined the impact of blogging on student motivation in the context of a mathematics course. Results indicated that students had favorable feelings towards the use of blogs, and even recommended their use in the other classes. \citet{Angelaina2012}, \citet{Marques2013},and \citet{Jara2012} also found that students were successfully engaged into blogging activities. \citet{Bartholomew2012} found that educational blogs encourages co-ownership and lateral communications, giving the students the opportunity to develop their own voice in the course.

The study conducted by \citet{RojasAlvarez2011} also used the participant observation method for collecting data, in addition to other methods based on self-reported data. This study examined the capacity of blogs to engage a group of secondary students in the learning of a foreign language. Participants were asked to post entries and communicate among them to accomplish academic activities and express their feelings and opinions. Results indicated that the blogging activity had fostered their writing habits, although not so much as expected. This partially positive results was attributed to the communication capabilities of blogs. 

Four studies followed an experimental design, i.e. with a control group. Three of them reported that students had positive perceptions towards the use of blogs. More specifically, \citet{Yang2012} conducted a experimental study in which both groups, control and treatment, perform blogging activities, but only the treatment group could make comments on colleague contributions. Results indicated that students of both groups were positively motivated to learn from peers, independently of the comment feature. \citet{RefWorks:137} and \citet{Chhabra2011} also found that the blogging activity had a positive effect in the students' perceptions. However, the study conducted by \citet{Hsu2011} showed that the use of blogs did not improve the students' learning motivation. 
%---------------------------------------------------------------------------------------------

%---------------------------------------------------------------------------------------------
\subsection{Effects of blogging on collaborative learning and students' interaction}

A common interest within the researcher community was to measure the impact of the collaborative and interactive features of blogs to support the collaborative learning and social construction of knowledge among students. Following the trend of the previous categories, almost all studies used data collecting methods based on participant self-reported data, such as surveys, interviews and content of the blogs. Besides, none of them had control group. Overall, results indicated that the commenting feature of blogs was successfully used to give comments and suggestions, and also to promote the interaction among participants (students and instructors). However, there was not collaboration for social construction of knowledge. Higher order competencies, such as coaching, were not acquired by the students either. These findings are derived from the following studies. \citet{RefWorks:139} conducted a study with 113 undergraduate students of Computer Science to examine the performance of a framework based on blogging activities to improve cognitive, social, and self-directed learning skills of students. Participants were asked to post entries in group blogs to reflect upon their laboratory practices, and also comment on classmates' blogs. Results indicated that students gave hints for solving problems, and encouraged their classmates by means of the commenting functionality of blogs. Nonetheless, no higher order skills, such as coaching, were achieved. On the other hand, blogging provided students with opportunities to develop self-directed learning behaviors. \citet{RefWorks:140} conducted a study with 24 undergraduate students of an ICT course to assess blogs as online tools that could foster the individual and collaborative learning, and the interaction among students. Participants were required to actively make contributions to personal blogs related to both informal and formal activities. Results indicated that blogs are an important tool to promote the interaction among students, and between the instructor and the students. The study also emphasized the decentralized nature of the instructor-student relationship that is achieved by the blog mediation, and the potential to share information and knowledge in a non-hierarchical way. \citet{RefWorks:145} examined the potential of blogs to develop evaluative skills through a peer-review process, and create a learning community in the context of journalism activities. The study found that most of the 52 students composed useful comments and suggestions. In addition, trusting relationships among groups of students were established, which led students to reflect on their work and take appropriate actions for their personal improvement. \citet{Maheridou2012} conducted a study with 20 physical educators to evaluate the students' interaction and the social construction of knowledge through the use of blogs. Participants were asked for creating group threaded discussions inside the blog with the classmates. Results showed a high level of students' interaction, and an active collaboration for knowledge construction by means of the sharing and comparison of information. \citet{Angelaina2012} analyzed the contents of a group blog containing the contributions of 21 secondary students to investigate their participation and learning presence. Students were required to contribute to an educational group blog with posts and comments to perform a cross-thematic inquiry task. Results suggested that an appropriate blog design can enhance the students' collaboration skills and the critical thinking. \citet{Alterman2013} conducted a study to analyze the knowledge creation, distribution, and accumulation among students via blogging activities. 25 undergraduate students from an Internet and Society course participated in the study, and were asked to write regular posts and commenting on each other's contributions. The study suggested that blogging activities can help students to create a basis of common knowledge to enhance the quality of the course assignments. \citet{Marques2013} examined the influence of quote and reply functionalities of blogs to promote the collaborative learning. 26 undergraduate participants were required to post information about the project and to comment the messages received from the classmates. Results indicated that blogs improved the learning of the subject via interaction and collaboration among students. \citet{Fessakis2013} conducted a study with 147 undergraduate students to assess the students' perceptions about the usability of blogs, and the promotion of communication and collaboration skills by blogging. Results indicated that most students considered blogs as easy and user-friendly tools, and they also recognized the suitability of blogs for communication and collaboration tasks. \citet{RefWorks:148} reported that students of an ICT course were able to discover new ways of learning practices without direct involvement of the instructor thanks to development of blogging activities with a peer-to-peer interaction. The study conducted by \citet{RefWorks:119} showed low levels of collaboration between the students, and suggested the need to increase students' participation. \citet{RefWorks:201} reported mixed results on the question if the use of blogs was an adequate mechanism for giving or receiving peer advice. The study conducted by \citet{RefWorks:147} found that the effects of blogs for developing extensive and dynamic dialogues were questionable, and that the interactive capability of blogs was mainly used to enhance the social presence and the socio-emotional dimension. \citet{Vurdien2013} reported that collaborative skills were strengthened by the students' interaction in blogs in the context of enhancement of writing skills of foreign languages. The study conducted by \citet{Garcia-Sanchez2012} also used the participant observation method to collect data, in addition to surveys and contents of blogs. They reported that the use of a bilingual blog can foster the students' participation and interaction thanks to the comment-based peer feedback.
%---------------------------------------------------------------------------------------------

%---------------------------------------------------------------------------------------------
\subsection{Effects of blogging on social support}

% IVÁN (16 feb 2013): Carlos, en el anterior estudio se indica que algunos estudiantes lo encontraron útil para ciertas cosas, mientras que otros lo encontraron más bien un consumo de tiempo innecesario. Si en el estudio original se indica cuantos estudiantes había de cada opinión indícalo en las anteriores frases. No es lo mismo que 80 estudiantes opinaran una cosa y 2 estudiantes otra, que 40 opinaran una cosa y 40 otra. Este comentario está dentro de un párrafo para que te sea más fácil ubicar la frase a la que me refiero.
% CARLOS R. (19 feb 2013): después de reelerme el artículo entero, no se aporta dicha información. Se repite la misma conclusión en diversas partes pero nunca da cifras o porcentajes.
% IVÁN (20 feb 2013): En ese caso, no te preocupes. Está bien como está. Si los autores originales del trabajo no lo mencionan, nosotros no podemos saberlo. Por tanto, lo dejamos como está.
Social and emotional support is another feature attributed to blogs that has called the attention of researchers. Most of the studies (10 out of 12) that compose this category are characterized by the lack of a control group. Also, ten out of twelve studies exclusively used self-reported data.

The overall results of the nine studies that did not have a control and were exclusively based on self-reported data indicated that students perceived the blogging activity as useful for venting emotions, and give social support between peers. Next, the studies supporting these findings are commented. \citet{RefWorks:201} interviewed 91 pre-service teachers by groups, who took a course about managing the learning environment for primary education. The study aimed at examining the effectiveness of blogs in order to gain real practice in the behavior management through reflective practices. Students were required to use individual blogs to share their experiences and reflect on the progress of a classroom plan that they had to develop. The study reported mixed results: whereas some students perceived blogging activity as useful for venting emotions, supporting each other, and acquiring advanced behavior management strategies, others considered the use of blogs as time consuming. \citet{RefWorks:147} interviewed and surveyed 37 student teachers to analyze the effectiveness of a developed framework that used blogs in the context of teacher education. Students were asked to write posts in personal blogs to reflect upon memorable incidents during teaching formation. They also had to read and comment on classmates' blogs to offer support and exchange of knowledge. Results indicated that blogs were valuable tools to support the emotionally charged and social-oriented individual expressions. Moreover, interactive functionality of blogs was mainly used to enhance the social presence and the socio-emotional dimension of the learning community. \citet{RefWorks:128} reported that blogs were mainly used for social and emotional support, showing the potential of blogs to create an atmosphere of beneficial social interaction that led to knowledge development. In a similar way, \citet{RefWorks:139} indicated that blogs allowed students to express their emotional experiences, and exchange social support with classmates. In addition, he argued that the social support was an important need for novice learners that were facing new challenges in their first year as undergraduate students. \citet{Pinilla2013} found that medical students used blogs to improve their social life beyond medical school and to deal with distressing situations. \citet{Alqudsi-Ghabra2012} reported that blogging extendeds the student's audience and allows shy students to voice their opinions. Other studies, such as \citet{RefWorks:119}, \citet{Maheridou2012} and \citet{RefWorks:125}, also found that students used blogs to provide one another with social and emotional support, which was essential in realizing meaningful educational outcomes. 

\citet{RefWorks:151} conducted an study that used participant observation, interviews, contents of the blogs as data collected methods. They argued that the improvement of the writing skills of 16 students of elementary education was attributed to a great extent to the support that the students received from their peers and the instructor through blogs. 

Two of the studies followed an experimental approach. The study conducted by \citet{Hsu2011} found a significant difference in the student retention rate in favor of the treatment group using blogs. This was attributed to higher interaction and social support among those students using blogs. \citet{Yang2012} conducted a experimental study to evaluate how blog comments can influence the students' perception towards peer interaction and peer learning. Both control and treatment groups were required to post entries in individual blogs, but only the treatment group could comment on classmates' posts. The study found that blog comments were positively perceived as a useful feature for peer interaction and academic achievements.
%---------------------------------------------------------------------------------------------

%---------------------------------------------------------------------------------------------
\subsection{Effects of blogging on the development of specific skills}

A number of studies have focused on examining the effects of blogs on specific skills, such as writing and ICT competencies. Six out of twelve studies exclusively relied on self-reported data, while the others six combined this kind of data with academic tests, assignments, and/or participant observation. On the other hand, only study had control group.

Regarding the writing literacy, the studies agreed that blogging activities improved the writing skills of students, and at the same time helped in developing their writer identities. For example, \citet{RefWorks:128} conducted a study to assess how the interactive and collaborative characteristics of blogs could help in the improvement of academic writing skills. Seven master students of linguistics and TESOL (Teachers of English to Speakers of Other Languages) were required to post entries in personal blogs to reflect upon specific issues of academic writing. Results indicated that blogging allowed students to improve their academic writing knowledge and develop their writer identities. \citet{RefWorks:145} interviewed by groups and surveyed 52 undergraduate students studying journalism with the purpose of evaluating whether the use of blogs could improve their writing skills, and create a sense of community for learning. Participants were required to post entries in group blogs about news-writing assignments. Obtained results showed that students developed proficiency in writing by means of the blogging activity and the feedback provided by the peers through comments. \citet{RefWorks:151} conducted a qualitative case study with 16 5th-grade elementary students to evaluate the effects of blogging in the area of writing literacy. Participant observation, interviews and contents of blogs were used as data collecting methods. Students were asked to post entries in individual blogs and reflect on how blogging helped them to develop their writing skills. The study concluded that students actually improved their writing skills, and took awareness of their writer identities. These results were attributed to the support the students received from the blogging community. \citet{Vurdien2013} examined whether blogs could enhance writing skills in specific writing tasks and foster the collaborative learning. Eleven participants were required to complete different writing tasks in personal blogs and make comments on peers' blogs. Results indicated that blogs can motivate students to improve their writing skills through self-reflection and peer feedback. \citet{Chen2012} conducted a study with 67 undergraduate students enrolled in an academic English Writing course to evaluate if the peer-review activities based on blogs could improve the students' writing skills of English as a foreign language. Results revealed the effectiveness of blog-based peer-reviewing as a means of improving the students' writing skills, and their perceptions about this process. \citet{Garcia-Sanchez2012} conducted a study 87 postgraduate students to examine if a bilingual blog could enhance their writing and reading skills in a foreign language, and also to promote the students' motivation and cooperation. Students were asked to post assignments and provide feedback about other colleague assignments in pairs. The study found that blog-based writing and peer revisions improve the students' writing and reading skills. The study conducted by \citet{HamadAljumah2011} also found an improvement in the writing skills of English for second language learners. The study conducted by \citet{Suryani2012} used surveys and academic tests as data collection methods. The study examined whether blogging based activities could enhance the writing skills in a foreign language using a sample of 37 undergraduate students. Participants were required to post writing assignments, make comments on colleagues' entries, and contribute in discussions initiated by instructors. Results indicated that students had improved their language competencies and had also enhanced their writing skills. \citet{Jara2012} conducted a study with primary students to evaluate whether blogging can enhance the writing skills of a foreign language, more specifically the use of English adjectives. Participants were required to do some exercises implemented in a blog, in addition to other worksheets in a hard copy form. The study found an improvement on the grammar and practical use of English adjectives.

The researchers that addressed the issue of whether blog usage was potentially beneficial to develop ICT skills reported the positive convenience of blogs for this purpose. For instance, \citet{RefWorks:123} interviewed 339 prospective teachers with the purpose of examining the effects of blogging on the perception of ICT competencies. Using group blogs, participants had to discuss issues about an assigned topic. The study found that participants perceived an increase in their ICT skills by means of the active participation in the use of blogs. In addition, they positively changed their ICT perceptions, which can encourage prospective teachers to adopt blogs and other ICT tools in their future classrooms. \citet{RefWorks:132} argued that students who had taken ICT-related courses could be more inclined towards the use of ICT tools than other students who lacked such experiences. The study conducted by \citet{RefWorks:137}, the only one in this category that has control group, indicated that blogging activity had a positive impact on students' ICT skills.
%---------------------------------------------------------------------------------------------

%---------------------------------------------------------------------------------------------
\section{Discussion and conclusions}
\label{sec:conc}

Researchers have explored several dimensions of the use of blogs for educational purposes. Regarding the reported effects of blogging for developing reflective practices, which in turn could improve the student learning performance, the results are controversial. Most of the studies that lacked control group reported positive results, showing an improvement in the perceived reflective thinking of the students. This fact could allow students to take a step forward in their learning processes. However, the results of the studies that followed an experimental design present a very different trend. More than the half of them did not find a significant learning improvement in the treatment group (the one that had used blogs) in comparison with the control group (the one that had not used blogs). Although, it is true that number of non-experimental studies (24 out of 29) is much greater than the experimental ones, the results of the experimental studies are subject to a less bias, and therefore they present more robust and conclusive results. For this reason, the different trend in the results between experimental and non experimental studies is so meaningful. The conclusion is that more research has to be performed to either categorically claim the benefits of blogs in support of reflective thinking and students' learning, or on the contrary discard the use of blogs as a useful tool for learning and teaching. On the other hand, it should be investigated the conjecture of \citet{RefWorks:133} that hypothesized that one possible reason of not having found positive results could be the lack of direct instruction for performing blogging activities. He worked under the assumption that students had a sounded knowledge and a high experience with Web 2.0 tools. In the studies under review, this issue is not always explicitly documented, and consequently it is no possible to measure its impact in the reported results.

On the other hand, \citet{RefWorks:101} recommended focusing on the research of situations and strategies in which the blogs could be best used as an instructional tool, rather than examining the effects of blogs in the students' learning. They argued that the comparison of different media (in this case blogs versus no blogs) could be especially problematic due to the existence of many potential confounding variables. However, it is worth mentioning that it is important to know the true learning and teaching capabilities of blogs (or any other technological tool) before introducing them massively in the instructional process. Thus, the current review recommends clarifying the performance of the blogs for learning and teaching purposes.

Regarding the effects of blogs on the perceived student engagement in learning tasks, the results indicated that students showed positive feelings towards the use of blogs and a higher interest in the learning activities. However, except one study conducted by \citet{RefWorks:137}, all of the studies relied on self-report data, and lacked of a control group. Therefore, the conduction of new research studies with an experimental design is recommended to corroborate these results.

The conclusions regarding the effects of blogs on the collaborative learning and social construction of knowledge are that students successfully wrote comments to give tips and suggestions, which also increased the interaction among the participants. Nonetheless, coaching or other high levels of collaboration were not found for construction of knowledge. Again, the majority of studies relied on participant self-reported data, such as surveys, interviews and contents of the blogs. In any case, with the purpose of increase the students' participation and collaboration, the adoption of measures to ensure that students have an appropriate level of understanding and familiarity with blogs is highly recommended. Some studies have showed that students who are more skillful with the blogging technology participated more often, and even their contributions were more reflective and valuable~\citep{RefWorks:134}. Therefore, the strategy to follow is to offer an explicit guidance to increase the ability and confidence of the students, which in turn would cause a higher participation in the blogging activity. This increase of participation would not only gradually improve the quality of the contributions, but also would shorten the time for elaborating the postings, solving at the same time some frequent negative perceptions related to the time that students have to spend in blogging.

With regard to the capabilities of blogs to give social and emotional support, the findings indicate that students have positive perceptions on the use of blogs to vent emotions and give social support between peers. The corresponding studies were also characterized by the use of self-reported data.

The reviewed studies agree on that the use of blogs can increase the mastery of specific skills, such as writing literacy, academic writing, and those related with the acquisition of ICT competencies. 

An additional issue that must be taken into account is general typology of the reviewed studies. Most of them are based exclusively on self-reported data (33 out of 47). These studies present some limitations since the participants tend to give answers that cause them to be highly esteemed~\citep{RefWorks:204,RefWorks:205}. This usually occurs because of their preconceived notions about which the socially desirable answers are. Another issue that can affect the confidence and reliability of the reported results of the reviewed studies is the lack of an experimental design, i.e. the lack of a control group. The number of studies with control group is only six, which is very low in comparison with the studies without control group (41 studies). Therefore, it is highly recommended that the future studies follow an experimental design and use a combination of data collection methods not only based on self-reported data with the purpose of obtaining more robust and conclusive results.

In comparison with the review conducted by \citet{RefWorks:101}, there has been an increase in the number of studies that have followed an experimental design: from one to six studies. Nonetheless, this quantity is still not enough in absolute terms to properly assess the effects of blogs.
%---------------------------------------------------------------------------------------------

%---------------------------------------------------------------------------------------------
\subsubsection*{Acknowledgments}
%---------------------------------------------------------------------------------------------

%---------------------------------------------------------------------------------------------
\bibliographystyle{model5-names}
\bibliography{blogs}

\begin{thebibliography}{68}
\expandafter\ifx\csname natexlab\endcsname\relax\def\natexlab#1{#1}\fi
\providecommand{\url}[1]{\texttt{#1}}
\providecommand{\href}[2]{#2}
\providecommand{\path}[1]{#1}
\providecommand{\DOIprefix}{doi:}
\providecommand{\ArXivprefix}{arXiv:}
\providecommand{\URLprefix}{URL: }
\providecommand{\Pubmedprefix}{pmid:}
\providecommand{\doi}[1]{\href{http://dx.doi.org/#1}{\path{#1}}}
\providecommand{\Pubmed}[1]{\href{pmid:#1}{\path{#1}}}
\providecommand{\bibinfo}[2]{#2}
\ifx\xfnm\relax \def\xfnm[#1]{\unskip,\space#1}\fi
%Type = Article
\bibitem[{Alqudsi-Ghabra \& Al-Bahrani(2012)}]{Alqudsi-Ghabra2012}
\bibinfo{author}{Alqudsi-Ghabra, T.}, \& \bibinfo{author}{Al-Bahrani, M.}
  (\bibinfo{year}{2012}).
\newblock \bibinfo{title}{{Educational Blogging: The Case of Graduate MLIS
  Students in Kuwait}}.
\newblock {\it \bibinfo{journal}{International Journal of Libraries and
  Information Services}\/},  {\it \bibinfo{volume}{62}\/},
  \bibinfo{pages}{305--402}.
%Type = Article
\bibitem[{Alterman \& Larusson(2013)}]{Alterman2013}
\bibinfo{author}{Alterman, R.}, \& \bibinfo{author}{Larusson, J.}
  (\bibinfo{year}{2013}).
\newblock \bibinfo{title}{{Participation and common knowledge in a case study
  of student blogging}}.
\newblock {\it \bibinfo{journal}{International Journal of Computer-Supported
  Collaborative Learning}\/},  {\it \bibinfo{volume}{8}\/},
  \bibinfo{pages}{149--187}.
%Type = Article
\bibitem[{Angelaina \& Jimoyiannis(2012)}]{Angelaina2012}
\bibinfo{author}{Angelaina, S.}, \& \bibinfo{author}{Jimoyiannis, A.}
  (\bibinfo{year}{2012}).
\newblock \bibinfo{title}{{Analysing students' engagement and learning presence
  in an educational blog community}}.
\newblock {\it \bibinfo{journal}{Educational Media International}\/},  {\it
  \bibinfo{volume}{49}\/}, \bibinfo{pages}{183--200}.
%Type = Book
\bibitem[{Argirys \& Schön(1978)}]{RefWorks:194}
\bibinfo{author}{Argirys, C.}, \& \bibinfo{author}{Schön, D.~A.}
  (\bibinfo{year}{1978}).
\newblock {\it \bibinfo{title}{{Organizational learning: A theory of action
  perspective}}\/}.
\newblock \bibinfo{publisher}{Addison-Wesley}.
%Type = Article
\bibitem[{Askim et~al.(2011)Askim, IZMIRLI \& SAHIN-IZMIRLI}]{RefWorks:141}
\bibinfo{author}{Askim, K.~A.}, \bibinfo{author}{IZMIRLI, S.}, \&
  \bibinfo{author}{SAHIN-IZMIRLI, O.} (\bibinfo{year}{2011}).
\newblock \bibinfo{title}{{Student Experience in Blog Use for Supplementary
  Purposes in Courses}}.
\newblock {\it \bibinfo{journal}{Turkish Online Journal of Distance
  Education}\/},  {\it \bibinfo{volume}{12}\/}, \bibinfo{pages}{78--96}.
%Type = Article
\bibitem[{Bartholomew et~al.(2012)Bartholomew, Jones \&
  Glassman}]{Bartholomew2012}
\bibinfo{author}{Bartholomew, M.}, \bibinfo{author}{Jones, T.}, \&
  \bibinfo{author}{Glassman, M.} (\bibinfo{year}{2012}).
\newblock \bibinfo{title}{{A Community of Voices: Educational Blog Management
  Strategies and Tools}}.
\newblock {\it \bibinfo{journal}{TechTrends}\/},  {\it \bibinfo{volume}{56}\/},
  \bibinfo{pages}{19--25}.
%Type = Article
\bibitem[{Boud \& Walker(1998)}]{RefWorks:202}
\bibinfo{author}{Boud, D.}, \& \bibinfo{author}{Walker, D.}
  (\bibinfo{year}{1998}).
\newblock \bibinfo{title}{Promoting reflection in professional courses: The
  challenge of context}.
\newblock {\it \bibinfo{journal}{Studies in higher education}\/},  {\it
  \bibinfo{volume}{23}\/}, \bibinfo{pages}{191--206}.
%Type = Article
\bibitem[{Brookfield(1998)}]{RefWorks:199}
\bibinfo{author}{Brookfield, S.} (\bibinfo{year}{1998}).
\newblock \bibinfo{title}{Critically reflective practice}.
\newblock {\it \bibinfo{journal}{Journal of Continuing Education in the Health
  Professions}\/},  {\it \bibinfo{volume}{18}\/}, \bibinfo{pages}{197--205}.
%Type = Article
\bibitem[{Cain \& Giraud(2012)}]{Cain2012}
\bibinfo{author}{Cain, H.}, \& \bibinfo{author}{Giraud, V.}
  (\bibinfo{year}{2012}).
\newblock \bibinfo{title}{{Critical Thinking Skills Evidenced in Graduate
  Students Blogs}}.
\newblock {\it \bibinfo{journal}{Journal of Leadership Education}\/},  {\it
  \bibinfo{volume}{11}\/}, \bibinfo{pages}{72--87}.
%Type = Article
\bibitem[{Cameron(2012)}]{RefWorks:134}
\bibinfo{author}{Cameron, M.~P.} (\bibinfo{year}{2012}).
\newblock \bibinfo{title}{{`Economics with Training Wheels': Using Blogs in
  Teaching and Assessing Introductory Economics}}.
\newblock {\it \bibinfo{journal}{Journal of Economic Education}\/},  {\it
  \bibinfo{volume}{43}\/}, \bibinfo{pages}{397--407}.
%Type = Article
\bibitem[{Chen(2012)}]{Chen2012}
\bibinfo{author}{Chen, K. T.-C.} (\bibinfo{year}{2012}).
\newblock \bibinfo{title}{{Blog-Based Peer Reviewing in EFL Writing Classrooms
  for Chinese Speakers}}.
\newblock {\it \bibinfo{journal}{Computers and Composition}\/},  {\it
  \bibinfo{volume}{29}\/}, \bibinfo{pages}{280--291}.
%Type = Article
\bibitem[{Chhabra \& Sharma(2011)}]{Chhabra2011}
\bibinfo{author}{Chhabra, R.}, \& \bibinfo{author}{Sharma, V.}
  (\bibinfo{year}{2011}).
\newblock \bibinfo{title}{{Applications of blogging in problem based
  learning}}.
\newblock {\it \bibinfo{journal}{Education and Information Technologies}\/},
  {\it \bibinfo{volume}{18}\/}, \bibinfo{pages}{3--13}.
%Type = Article
\bibitem[{Chu et~al.(2012)Chu, Chan \& Tiwari}]{RefWorks:125}
\bibinfo{author}{Chu, S.~K.}, \bibinfo{author}{Chan, C.~K.}, \&
  \bibinfo{author}{Tiwari, A.~F.} (\bibinfo{year}{2012}).
\newblock \bibinfo{title}{{Using blogs to support learning during internship}}.
\newblock {\it \bibinfo{journal}{Computers \& Education}\/},  {\it
  \bibinfo{volume}{58}\/}, \bibinfo{pages}{989--1000}.
%Type = Article
\bibitem[{Churchill(2011)}]{RefWorks:150}
\bibinfo{author}{Churchill, D.} (\bibinfo{year}{2011}).
\newblock \bibinfo{title}{{Web 2.0 in Education: A Study of the Explorative Use
  of Blogs with a Postgraduate Class}}.
\newblock {\it \bibinfo{journal}{Innovations in Education and Teaching
  International}\/},  {\it \bibinfo{volume}{48}\/}, \bibinfo{pages}{149--158}.
%Type = Article
\bibitem[{Deed \& Edwards(2011)}]{Deed2011}
\bibinfo{author}{Deed, C.}, \& \bibinfo{author}{Edwards, a.}
  (\bibinfo{year}{2011}).
\newblock \bibinfo{title}{{Unrestricted student blogging: Implications for
  active learning in a virtual text-based environment}}.
\newblock {\it \bibinfo{journal}{Active Learning in Higher Education}\/},  {\it
  \bibinfo{volume}{12}\/}, \bibinfo{pages}{11--21}.
%Type = Article
\bibitem[{Deng \& Yuen(2011)}]{RefWorks:147}
\bibinfo{author}{Deng, L.}, \& \bibinfo{author}{Yuen, A. H.~K.}
  (\bibinfo{year}{2011}).
\newblock \bibinfo{title}{{Towards a Framework for Educational Affordances of
  Blogs}}.
\newblock {\it \bibinfo{journal}{Computers \& Education}\/},  {\it
  \bibinfo{volume}{56}\/}, \bibinfo{pages}{441--451}.
%Type = Article
\bibitem[{Fessakis et~al.(2013)Fessakis, Dimitracopoulou \&
  Palaiodimos}]{Fessakis2013}
\bibinfo{author}{Fessakis, G.}, \bibinfo{author}{Dimitracopoulou, A.}, \&
  \bibinfo{author}{Palaiodimos, A.} (\bibinfo{year}{2013}).
\newblock \bibinfo{title}{{Graphical Interaction Analysis Impact on Groups
  Collaborating through Blogs}}.
\newblock {\it \bibinfo{journal}{Educational Technology \& Society}\/},  {\it
  \bibinfo{volume}{16}\/}, \bibinfo{pages}{243--253}.
%Type = Article
\bibitem[{García-Sabater et~al.(2011)García-Sabater, Vidal-Carreras, Santandreu
  \& Perello-Marin}]{Garcia-Sabater2011}
\bibinfo{author}{García-Sabater, J.~J.}, \bibinfo{author}{Vidal-Carreras,
  P.~I.}, \bibinfo{author}{Santandreu, C.}, \& \bibinfo{author}{Perello-Marin,
  R.} (\bibinfo{year}{2011}).
\newblock \bibinfo{title}{{Practical experience in teaching inventory
  management with Edublogs}}.
\newblock {\it \bibinfo{journal}{Journal of Industrial Engineering and
  Management}\/},  {\it \bibinfo{volume}{4}\/}, \bibinfo{pages}{103--122}.
%Type = Article
\bibitem[{Garc\'{\i}a-S\'{a}nchez \& Rojas-Lizana(2012)}]{Garcia-Sanchez2012}
\bibinfo{author}{Garc\'{\i}a-S\'{a}nchez, S.}, \&
  \bibinfo{author}{Rojas-Lizana, S.} (\bibinfo{year}{2012}).
\newblock \bibinfo{title}{{Bridging the language and cultural gaps: the use of
  blogs}}.
\newblock {\it \bibinfo{journal}{Technology, Pedagogy and Education}\/},  {\it
  \bibinfo{volume}{21}\/}, \bibinfo{pages}{37--41}.
%Type = Book
\bibitem[{Gibbs et~al.(1988)Gibbs, Britain \& Unit}]{RefWorks:197}
\bibinfo{author}{Gibbs, G.}, \bibinfo{author}{Britain, G.}, \&
  \bibinfo{author}{Unit, F.~E.} (\bibinfo{year}{1988}).
\newblock {\it \bibinfo{title}{Learning by doing: a guide to teaching and
  learning methods}\/}.
\newblock \bibinfo{publisher}{Further Education Unit}.
%Type = Book
\bibitem[{Glaser \& Strauss(1967)}]{RefWorks:191}
\bibinfo{author}{Glaser, B.}, \& \bibinfo{author}{Strauss, A.}
  (\bibinfo{year}{1967}).
\newblock {\it \bibinfo{title}{{The Discovery of Grounded Theory: Strategies
  for Qualitative Research}}\/}.
\newblock \bibinfo{publisher}{Aldine Transaction}.
%Type = Article
\bibitem[{Goktas \& Demirel(2012)}]{RefWorks:123}
\bibinfo{author}{Goktas, Y.}, \& \bibinfo{author}{Demirel, T.}
  (\bibinfo{year}{2012}).
\newblock \bibinfo{title}{{Blog-enhanced ICT courses: Examining their effects
  on prospective teachers' ICT competencies and perceptions}}.
\newblock {\it \bibinfo{journal}{Computers \& Education}\/},  {\it
  \bibinfo{volume}{58}\/}, \bibinfo{pages}{908--917}.
%Type = Article
\bibitem[{Grant \& Booth(2009)}]{RefWorks:211}
\bibinfo{author}{Grant, M.~J.}, \& \bibinfo{author}{Booth, A.}
  (\bibinfo{year}{2009}).
\newblock \bibinfo{title}{A typology of reviews: an analysis of 14 review types
  and associated methodologies}.
\newblock {\it \bibinfo{journal}{Health Information \& Libraries Journal}\/},
  {\it \bibinfo{volume}{26}\/}, \bibinfo{pages}{91--108}.
%Type = Article
\bibitem[{Hakkarainen et~al.(2001)Hakkarainen, Muukonen, Lipponen, Ilomäki,
  Rahikainen \& Lehtinen}]{RefWorks:204}
\bibinfo{author}{Hakkarainen, K.}, \bibinfo{author}{Muukonen, H.},
  \bibinfo{author}{Lipponen, L.}, \bibinfo{author}{Ilomäki, L.},
  \bibinfo{author}{Rahikainen, M.}, \& \bibinfo{author}{Lehtinen, E.}
  (\bibinfo{year}{2001}).
\newblock \bibinfo{title}{{Teachers' information and communication technology
  (ICT) skills and practices of using ICT}}.
\newblock {\it \bibinfo{journal}{Journal of Technology and Teacher
  Education}\/},  {\it \bibinfo{volume}{9}\/}, \bibinfo{pages}{181--197}.
%Type = Article
\bibitem[{{Hamad Aljumah}(2011)}]{HamadAljumah2011}
\bibinfo{author}{{Hamad Aljumah}, F.} (\bibinfo{year}{2011}).
\newblock \bibinfo{title}{{Saudi Learner Perceptions and Attitudes towards the
  Use of Blogs in Teaching English Writing Course for EFL Majors at Qassim
  University}}.
\newblock {\it \bibinfo{journal}{English Language Teaching}\/},  {\it
  \bibinfo{volume}{5}\/}, \bibinfo{pages}{100--116}.
%Type = Article
\bibitem[{Hancock \& Flowers(2001)}]{RefWorks:205}
\bibinfo{author}{Hancock, D.~R.}, \& \bibinfo{author}{Flowers, C.~P.}
  (\bibinfo{year}{2001}).
\newblock \bibinfo{title}{{Comparing social desirability responding on World
  Wide Web and paper-administered surveys}}.
\newblock {\it \bibinfo{journal}{Educational technology research and
  development}\/},  {\it \bibinfo{volume}{49}\/}, \bibinfo{pages}{5--13}.
%Type = Article
\bibitem[{Harland \& Wondra(2011)}]{RefWorks:142}
\bibinfo{author}{Harland, D.~J.}, \& \bibinfo{author}{Wondra, J.~D.}
  (\bibinfo{year}{2011}).
\newblock \bibinfo{title}{{Preservice Teachers' Reflection on Clinical
  Experiences: A Comparison of Blog and Final Paper Assignments}}.
\newblock {\it \bibinfo{journal}{Journal of Digital Learning in Teacher
  Education}\/},  {\it \bibinfo{volume}{27}\/}, \bibinfo{pages}{128--133}.
%Type = Inproceedings
\bibitem[{Herring et~al.(2004)Herring, Scheidt, Bonus \& Wright}]{RefWorks:115}
\bibinfo{author}{Herring, S.~C.}, \bibinfo{author}{Scheidt, L.~A.},
  \bibinfo{author}{Bonus, S.}, \& \bibinfo{author}{Wright, E.}
  (\bibinfo{year}{2004}).
\newblock \bibinfo{title}{{Bridging the gap: A genre analysis of weblogs}}.
\newblock In {\it \bibinfo{booktitle}{Proc. Annual Hawaii International
  Conference on System Sciences}\/} (pp. \bibinfo{pages}{1--11}).
\newblock \bibinfo{publisher}{IEEE}.
%Type = Article
\bibitem[{Hodgson \& Wong(2011)}]{RefWorks:145}
\bibinfo{author}{Hodgson, P.}, \& \bibinfo{author}{Wong, D.}
  (\bibinfo{year}{2011}).
\newblock \bibinfo{title}{{Developing Professional Skills in Journalism through
  Blogs}}.
\newblock {\it \bibinfo{journal}{Assessment \& Evaluation in Higher
  Education}\/},  {\it \bibinfo{volume}{36}\/}, \bibinfo{pages}{197--211}.
%Type = Article
\bibitem[{Hsu \& Wang(2011)}]{Hsu2011}
\bibinfo{author}{Hsu, H.}, \& \bibinfo{author}{Wang, S.}
  (\bibinfo{year}{2011}).
\newblock \bibinfo{title}{{The impact of using blogs on college students'
  reading comprehension and learning motivation}}.
\newblock {\it \bibinfo{journal}{Literacy Research and Instruction}\/},  {\it
  \bibinfo{volume}{50}\/}, \bibinfo{pages}{68--88}.
%Type = Article
\bibitem[{Huffaker(2005)}]{RefWorks:206}
\bibinfo{author}{Huffaker, D.} (\bibinfo{year}{2005}).
\newblock \bibinfo{title}{{The educated blogger: Using weblogs to promote
  literacy in the classroom}}.
\newblock {\it \bibinfo{journal}{AACE Journal}\/},  {\it
  \bibinfo{volume}{13}\/}, \bibinfo{pages}{91--98}.
%Type = Article
\bibitem[{Hume(2012)}]{Hume2012}
\bibinfo{author}{Hume, M.} (\bibinfo{year}{2012}).
\newblock \bibinfo{title}{{Adopting organisation learning theory in the
  classroom: advancing learning through the use of blogging and
  self-reflection}}.
\newblock {\it \bibinfo{journal}{International Journal of Learning and
  Change}\/},  {\it \bibinfo{volume}{6}\/}, \bibinfo{pages}{49--65}.
%Type = Article
\bibitem[{Hungerford-Kresser et~al.(2012)Hungerford-Kresser, Wiggins \&
  Amaro-Jiménez}]{RefWorks:133}
\bibinfo{author}{Hungerford-Kresser, H.}, \bibinfo{author}{Wiggins, J.}, \&
  \bibinfo{author}{Amaro-Jiménez, C.} (\bibinfo{year}{2012}).
\newblock \bibinfo{title}{{Learning from Our Mistakes: What Matters When
  Incorporating Blogging in the Content Area Literacy Classroom}}.
\newblock {\it \bibinfo{journal}{Journal of Adolescent \& Adult Literacy}\/},
  {\it \bibinfo{volume}{55}\/}, \bibinfo{pages}{326--335}.
%Type = Article
\bibitem[{Jara(2012)}]{Jara2012}
\bibinfo{author}{Jara, O.~L.} (\bibinfo{year}{2012}).
\newblock \bibinfo{title}{{Using a Blog to Guide Beginner Students to Use
  Adjectives Appropriately When Writing Descriptions in English}}.
\newblock {\it \bibinfo{journal}{Profile Issues in Teachers' Professional
  Development}\/},  {\it \bibinfo{volume}{14}\/}, \bibinfo{pages}{187--209}.
%Type = Article
\bibitem[{Jimoyiannis(2012)}]{RefWorks:119}
\bibinfo{author}{Jimoyiannis, A.} (\bibinfo{year}{2012}).
\newblock \bibinfo{title}{{Towards an Analysis Framework for Investigating
  Students' Engagement and Learning in Educational Blogs}}.
\newblock {\it \bibinfo{journal}{Journal of Computer Assisted Learning}\/},
  {\it \bibinfo{volume}{28}\/}, \bibinfo{pages}{222--234}.
%Type = Article
\bibitem[{Johns(1995)}]{RefWorks:198}
\bibinfo{author}{Johns, C.} (\bibinfo{year}{1995}).
\newblock \bibinfo{title}{{Framing learning through reflection within Carper's
  fundamental ways of knowing in nursing}}.
\newblock {\it \bibinfo{journal}{Journal of advanced nursing}\/},  {\it
  \bibinfo{volume}{22}\/}, \bibinfo{pages}{226--234}.
%Type = Article
\bibitem[{Kang et~al.(2011)Kang, Bonk \& Kim}]{RefWorks:140}
\bibinfo{author}{Kang, I.}, \bibinfo{author}{Bonk, C.~J.}, \&
  \bibinfo{author}{Kim, M.-C.} (\bibinfo{year}{2011}).
\newblock \bibinfo{title}{{A Case Study of Blog-Based Learning in Korea:
  Technology becomes Pedagogy}}.
\newblock {\it \bibinfo{journal}{Internet and Higher Education}\/},  {\it
  \bibinfo{volume}{14}\/}, \bibinfo{pages}{227--235}.
%Type = Book
\bibitem[{Kolb(1984)}]{RefWorks:195}
\bibinfo{author}{Kolb, D.~A.} (\bibinfo{year}{1984}).
\newblock {\it \bibinfo{title}{{Experiential learning: Experience as the source
  of learning and development}}\/}.
\newblock \bibinfo{publisher}{Prentice Hall}.
%Type = Article
\bibitem[{Liu et~al.(2012)Liu, Kalk, Kinney \& Orr}]{RefWorks:212}
\bibinfo{author}{Liu, M.}, \bibinfo{author}{Kalk, D.}, \bibinfo{author}{Kinney,
  L.}, \& \bibinfo{author}{Orr, G.} (\bibinfo{year}{2012}).
\newblock \bibinfo{title}{{Web 2.0 and Its Use in Higher Education from
  2007-2009: A Review of Literature}}.
\newblock {\it \bibinfo{journal}{International Journal on E-Learning}\/},  {\it
  \bibinfo{volume}{11}\/}, \bibinfo{pages}{153--179}.
%Type = Article
\bibitem[{Maheridou et~al.(2012)Maheridou, Antoniou, Tsitskari \&
  Kourtessis}]{Maheridou2012}
\bibinfo{author}{Maheridou, M.}, \bibinfo{author}{Antoniou, P.},
  \bibinfo{author}{Tsitskari, E.}, \& \bibinfo{author}{Kourtessis, T.}
  (\bibinfo{year}{2012}).
\newblock \bibinfo{title}{{Network and Content Analysis in a Blog Training
  Course}}.
\newblock {\it \bibinfo{journal}{International J. Soc. Sci. \& Education}\/},
  {\it \bibinfo{volume}{2}\/}, \bibinfo{pages}{238--251}.
%Type = Article
\bibitem[{Manfra \& Lee(2012)}]{Manfra2012}
\bibinfo{author}{Manfra, M.~M.}, \& \bibinfo{author}{Lee, J.~K.}
  (\bibinfo{year}{2012}).
\newblock \bibinfo{title}{{You have to know the past to (blog) the present:
  Using an Educational Blog to Engage Students in U.S. History}}.
\newblock {\it \bibinfo{journal}{Computers in the Schools}\/},  {\it
  \bibinfo{volume}{29}\/}, \bibinfo{pages}{118--134}.
%Type = Article
\bibitem[{Marques et~al.(2013)Marques, Krejci, Siqueira, Pimentel \&
  Braz}]{Marques2013}
\bibinfo{author}{Marques, A.~M.}, \bibinfo{author}{Krejci, R.},
  \bibinfo{author}{Siqueira, S.~W.}, \bibinfo{author}{Pimentel, M.}, \&
  \bibinfo{author}{Braz, M. H.~L.} (\bibinfo{year}{2013}).
\newblock \bibinfo{title}{{Structuring the discourse on social networks for
  learning: Case studies on blogs and microblogs}}.
\newblock {\it \bibinfo{journal}{Computers in Human Behavior}\/},  {\it
  \bibinfo{volume}{29}\/}, \bibinfo{pages}{395--400}.
%Type = Article
\bibitem[{McGrail \& Davis(2011)}]{RefWorks:151}
\bibinfo{author}{McGrail, E.}, \& \bibinfo{author}{Davis, A.}
  (\bibinfo{year}{2011}).
\newblock \bibinfo{title}{{The Influence of Classroom Blogging on Elementary
  Student Writing}}.
\newblock {\it \bibinfo{journal}{Journal of Research in Childhood
  Education}\/},  {\it \bibinfo{volume}{25}\/}, \bibinfo{pages}{415--437}.
%Type = Article
\bibitem[{Minocha(2009)}]{RefWorks:214}
\bibinfo{author}{Minocha, S.} (\bibinfo{year}{2009}).
\newblock \bibinfo{title}{{Role of Social Software Tools in Education: A
  Literature Review}}.
\newblock {\it \bibinfo{journal}{Education \& Training}\/},  {\it
  \bibinfo{volume}{51}\/}, \bibinfo{pages}{353--369}.
%Type = Article
\bibitem[{{Mohd Hashim}(2012)}]{MohdHashim2012}
\bibinfo{author}{{Mohd Hashim}, M.~H.} (\bibinfo{year}{2012}).
\newblock \bibinfo{title}{{Group Blogs as Toolkits to Support Learning
  Environments in Statistics Subject: A Qualitative Case Study}}.
\newblock {\it \bibinfo{journal}{International Education Studies}\/},  {\it
  \bibinfo{volume}{5}\/}, \bibinfo{pages}{199--204}.
%Type = Article
\bibitem[{Nehme(2011)}]{RefWorks:138}
\bibinfo{author}{Nehme, Z.} (\bibinfo{year}{2011}).
\newblock \bibinfo{title}{{Blogging and the Learning of Linear Algebra Concepts
  through Contextual Mathematics}}.
\newblock {\it \bibinfo{journal}{Mathematics Teaching}\/},  {\it
  \bibinfo{volume}{225}\/}, \bibinfo{pages}{43--48}.
%Type = Article
\bibitem[{O'Donnell(2006)}]{RefWorks:208}
\bibinfo{author}{O'Donnell, M.} (\bibinfo{year}{2006}).
\newblock \bibinfo{title}{{Blogging as pedagogic practice: Artefact and
  ecology}}.
\newblock {\it \bibinfo{journal}{Asia Pacific Media Educator}\/},  {\it
  \bibinfo{volume}{1}\/}, \bibinfo{pages}{3}.
%Type = Article
\bibitem[{Olofsson et~al.(2011)Olofsson, Lindberg \& Hauge}]{RefWorks:148}
\bibinfo{author}{Olofsson, A.~D.}, \bibinfo{author}{Lindberg, J.~O.}, \&
  \bibinfo{author}{Hauge, T.~E.} (\bibinfo{year}{2011}).
\newblock \bibinfo{title}{{Blogs and the Design of Reflective Peer-to-Peer
  Technology-Enhanced Learning and Formative Assessment}}.
\newblock {\it \bibinfo{journal}{Campus-Wide Information Systems}\/},  {\it
  \bibinfo{volume}{28}\/}, \bibinfo{pages}{183--194}.
%Type = Article
\bibitem[{Osman \& Koh(2013)}]{Osman2013}
\bibinfo{author}{Osman, G.}, \& \bibinfo{author}{Koh, J. H.~L.}
  (\bibinfo{year}{2013}).
\newblock \bibinfo{title}{{Understanding management students' reflective
  practice through blogging}}.
\newblock {\it \bibinfo{journal}{The Internet and Higher Education}\/},  {\it
  \bibinfo{volume}{16}\/}, \bibinfo{pages}{23--31}.
%Type = Article
\bibitem[{Papastergiou et~al.(2011)Papastergiou, Gerodimos \&
  Antoniou}]{RefWorks:137}
\bibinfo{author}{Papastergiou, M.}, \bibinfo{author}{Gerodimos, V.}, \&
  \bibinfo{author}{Antoniou, P.} (\bibinfo{year}{2011}).
\newblock \bibinfo{title}{{Multimedia Blogging in Physical Education: Effects
  on Student Knowledge and ICT Self-Efficacy}}.
\newblock {\it \bibinfo{journal}{Computers \& Education}\/},  {\it
  \bibinfo{volume}{57}\/}, \bibinfo{pages}{1998--2010}.
%Type = Article
\bibitem[{Pinilla et~al.(2013)Pinilla, Weckbach, Alig, Bauer, Noerenberg,
  Singer \& Tiedt}]{Pinilla2013}
\bibinfo{author}{Pinilla, S.}, \bibinfo{author}{Weckbach, L.~T.},
  \bibinfo{author}{Alig, S.~K.}, \bibinfo{author}{Bauer, H.},
  \bibinfo{author}{Noerenberg, D.}, \bibinfo{author}{Singer, K.}, \&
  \bibinfo{author}{Tiedt, S.} (\bibinfo{year}{2013}).
\newblock \bibinfo{title}{{Blogging medical students: a qualitative analysis}}.
\newblock {\it \bibinfo{journal}{GMS Zeitschrift f\"{u}r medizinische
  Ausbildung}\/},  {\it \bibinfo{volume}{30}\/}.
%Type = Article
\bibitem[{Repman et~al.(2005)Repman, Zinskie \& Carlson}]{RefWorks:207}
\bibinfo{author}{Repman, J.}, \bibinfo{author}{Zinskie, C.}, \&
  \bibinfo{author}{Carlson, R.~D.} (\bibinfo{year}{2005}).
\newblock \bibinfo{title}{{Effective use of CMC tools in interactive online
  learning}}.
\newblock {\it \bibinfo{journal}{Computers in the Schools}\/},  {\it
  \bibinfo{volume}{22}\/}, \bibinfo{pages}{57--69}.
%Type = Article
\bibitem[{Reupert \& Dalgarno(2011)}]{RefWorks:201}
\bibinfo{author}{Reupert, A.}, \& \bibinfo{author}{Dalgarno, B.}
  (\bibinfo{year}{2011}).
\newblock \bibinfo{title}{{Using Online Blogs to Develop Student Teachers'
  Behaviour Management Approaches}}.
\newblock {\it \bibinfo{journal}{Australian Journal of Teacher Education}\/},
  {\it \bibinfo{volume}{36}\/}, \bibinfo{pages}{5}.
%Type = Article
\bibitem[{Robertson(2011)}]{RefWorks:139}
\bibinfo{author}{Robertson, J.} (\bibinfo{year}{2011}).
\newblock \bibinfo{title}{{The Educational Affordances of Blogs for
  Self-Directed Learning}}.
\newblock {\it \bibinfo{journal}{Computers \& Education}\/},  {\it
  \bibinfo{volume}{57}\/}, \bibinfo{pages}{1628--1644}.
%Type = Article
\bibitem[{{Rojas \'{A}lvarez}(2011)}]{RojasAlvarez2011}
\bibinfo{author}{{Rojas \'{A}lvarez}, G.} (\bibinfo{year}{2011}).
\newblock \bibinfo{title}{{Writing Using Blogs: A Way to Engage Colombian
  Adolescents in Meaningful Communication}}.
\newblock {\it \bibinfo{journal}{Profile Issues in TeachersProfessional
  Development}\/},  {\it \bibinfo{volume}{13}\/}, \bibinfo{pages}{11--27}.
%Type = Book
\bibitem[{Rolfe et~al.(2001)Rolfe, Freshwater \& Jasper}]{RefWorks:200}
\bibinfo{author}{Rolfe, G.}, \bibinfo{author}{Freshwater, D.}, \&
  \bibinfo{author}{Jasper, M.} (\bibinfo{year}{2001}).
\newblock {\it \bibinfo{title}{{Critical reflection for nursing and the helping
  professions: A user's guide}}\/}.
\newblock \bibinfo{publisher}{Palgrave Basingstoke}.
%Type = Book
\bibitem[{Rosenfeld et~al.(1996)Rosenfeld, Edwards, Thomas \&
  Booth-Kewley}]{RefWorks:210}
\bibinfo{author}{Rosenfeld, P.}, \bibinfo{author}{Edwards, J.~E.},
  \bibinfo{author}{Thomas, M.~D.}, \& \bibinfo{author}{Booth-Kewley, S.}
  (\bibinfo{year}{1996}).
\newblock {\it \bibinfo{title}{{How to conduct organizational surveys: A
  step-by-step guide}}\/}.
\newblock \bibinfo{publisher}{Sage Publications, Incorporated}.
%Type = Article
\bibitem[{Santoro et~al.(2011)Santoro, Caldarola \& Villella}]{RefWorks:100}
\bibinfo{author}{Santoro, E.}, \bibinfo{author}{Caldarola, P.}, \&
  \bibinfo{author}{Villella, A.} (\bibinfo{year}{2011}).
\newblock \bibinfo{title}{{Using Web 2.0 technologies and social media for the
  cardiologist's education and update}}.
\newblock {\it \bibinfo{journal}{Giornale italiano di cardiologia}\/},  {\it
  \bibinfo{volume}{12}\/}, \bibinfo{pages}{174--181}.
%Type = Article
\bibitem[{Sim \& Hew(2010)}]{RefWorks:101}
\bibinfo{author}{Sim, J. W.~S.}, \& \bibinfo{author}{Hew, K.~F.}
  (\bibinfo{year}{2010}).
\newblock \bibinfo{title}{{The use of weblogs in higher education settings: A
  review of empirical research}}.
\newblock {\it \bibinfo{journal}{Educational Research Review}\/},  {\it
  \bibinfo{volume}{5}\/}, \bibinfo{pages}{151--163}.
%Type = Article
\bibitem[{Stevens \& Brown(2011)}]{RefWorks:152}
\bibinfo{author}{Stevens, E.~Y.}, \& \bibinfo{author}{Brown, R.}
  (\bibinfo{year}{2011}).
\newblock \bibinfo{title}{{Lessons Learned from the Holocaust: Blogging to
  Teach Critical Multicultural Literacy}}.
\newblock {\it \bibinfo{journal}{Journal of Research on Technology in
  Education}\/},  {\it \bibinfo{volume}{44}\/}, \bibinfo{pages}{31--51}.
%Type = Article
\bibitem[{Sun \& Chang(2012)}]{RefWorks:128}
\bibinfo{author}{Sun, Y.}, \& \bibinfo{author}{Chang, Y.-j.}
  (\bibinfo{year}{2012}).
\newblock \bibinfo{title}{{Blogging to learn: Becoming EFL academic writers
  through collaborative dialogues}}.
\newblock {\it \bibinfo{journal}{Language Learning \& Technology}\/},  {\it
  \bibinfo{volume}{16}\/}, \bibinfo{pages}{43--61}.
%Type = Article
\bibitem[{Suryani et~al.(2012)Suryani, Hizwari, Islam \& Desa}]{Suryani2012}
\bibinfo{author}{Suryani, I.}, \bibinfo{author}{Hizwari, S.},
  \bibinfo{author}{Islam, M.~A.}, \& \bibinfo{author}{Desa, H.}
  (\bibinfo{year}{2012}).
\newblock \bibinfo{title}{{Using Weblog in Learning English and Encouraging
  Adaptation among International Students in Perlis}}.
\newblock {\it \bibinfo{journal}{Higher Education Studies}\/},  {\it
  \bibinfo{volume}{2}\/}, \bibinfo{pages}{27--31}.
%Type = Article
\bibitem[{Tang(2012)}]{Tang2012}
\bibinfo{author}{Tang, E.} (\bibinfo{year}{2012}).
\newblock \bibinfo{title}{{The Reflective Journey of Pre-service ESL Teachers:
  An Analysis of Interactive Blog Entries}}.
\newblock {\it \bibinfo{journal}{The Asia-Pacific Education Researcher}\/}, .
%Type = Article
\bibitem[{Top(2012)}]{RefWorks:132}
\bibinfo{author}{Top, E.} (\bibinfo{year}{2012}).
\newblock \bibinfo{title}{{Blogging as a Social Medium in Undergraduate
  Courses: Sense of Community Best Predictor of Perceived Learning}}.
\newblock {\it \bibinfo{journal}{Internet and Higher Education}\/},  {\it
  \bibinfo{volume}{15}\/}, \bibinfo{pages}{24--28}.
%Type = Article
\bibitem[{Vurdien(2013)}]{Vurdien2013}
\bibinfo{author}{Vurdien, R.} (\bibinfo{year}{2013}).
\newblock \bibinfo{title}{{Enhancing writing skills through blogging in an
  advanced English as a Foreign Language class in Spain}}.
\newblock {\it \bibinfo{journal}{Computer Assisted Language Learning}\/},  {\it
  \bibinfo{volume}{26}\/}, \bibinfo{pages}{126--143}.
%Type = Article
\bibitem[{Wood(2012)}]{RefWorks:135}
\bibinfo{author}{Wood, P.} (\bibinfo{year}{2012}).
\newblock \bibinfo{title}{{Blogs as Liminal Space: Student Teachers at the
  Threshold}}.
\newblock {\it \bibinfo{journal}{Technology, Pedagogy and Education}\/},  {\it
  \bibinfo{volume}{21}\/}, \bibinfo{pages}{85--99}.
%Type = Article
\bibitem[{Xie et~al.(2008)Xie, Ke \& Sharma}]{RefWorks:209}
\bibinfo{author}{Xie, Y.}, \bibinfo{author}{Ke, F.}, \&
  \bibinfo{author}{Sharma, P.} (\bibinfo{year}{2008}).
\newblock \bibinfo{title}{{The effect of peer feedback for blogging on college
  students' reflective learning processes}}.
\newblock {\it \bibinfo{journal}{The Internet and Higher Education}\/},  {\it
  \bibinfo{volume}{11}\/}, \bibinfo{pages}{18--25}.
%Type = Article
\bibitem[{Yang \& Chang(2012)}]{Yang2012}
\bibinfo{author}{Yang, C.}, \& \bibinfo{author}{Chang, Y.-S.}
  (\bibinfo{year}{2012}).
\newblock \bibinfo{title}{{Assessing the effects of interactive blogging on
  student attitudes towards peer interaction, learning motivation, and academic
  achievements}}.
\newblock {\it \bibinfo{journal}{Journal of Computer Assisted Learning}\/},
  {\it \bibinfo{volume}{28}\/}, \bibinfo{pages}{126--135}.

\end{thebibliography}
%---------------------------------------------------------------------------------------------

\end{document}